\documentstyle[11pt]{article}
\renewcommand{\mathrm}[1]{{\rm #1}}

\renewcommand{\theequation}{\arabic{section}.\arabic{equation}}

\begin{document}
\baselineskip 19pt
\parskip 7pt
\thispagestyle{empty}


\vspace{24pt}

\begin{center}
{\large\bf Yangian Symmetry and 
Quantum Inverse Scattering Method for the One-Dimensional Hubbard 
Model
}

\vspace{24pt}

Shuichi {\sc Murakami}\ ${}^{1}$
\footnote[3]{E-mail:{\tt murakami@appi.t.u-tokyo.ac.jp}} 
and Frank {\sc G\"ohmann}\ ${}^{2}$
\footnote[4]{E-mail:{\tt frank@monet.phys.s.u-tokyo.ac.jp}}

\vspace{8pt}
${ }^{1}$  {\sl Department 
of Applied Physics, Faculty of Engineering,} \\
{\sl University of Tokyo,} \\
{\sl Hongo 7-3-1, Bunkyo-ku, Tokyo 113, Japan.} \\
${ }^{2}$ {\sl Department of Physics, Faculty of Science,} \\
{\sl University of Tokyo,} \\
{\sl Hongo 7-3-1, Bunkyo-ku, Tokyo 113, Japan.}\footnote[5]{ 
Address from Oct.\ 1996: Physikalisches Institut der Universit\"at
Bayreuth, TP1, 95440 Bayreuth, Germany.}

(Received:\makebox[15em]{} 

\end{center}

\vspace{6pt}

\begin{center}
{\bf Abstract}
\end{center}

We develop the quantum inverse scattering method for the
one-dimensional Hubbard model on the infinite interval at zero
density. $R$-matrix and monodromy matrix are obtained as limits
from their known counterparts on the finite interval.
The $R$-matrix greatly simplifies in the considered
limit. The new $R$-matrix contains a submatrix which turns into
the rational $R$-matrix of the XXX-chain by an appropriate
reparametrization. The corresponding submatrix of the monodromy
matrix thus provides a representation of the Y(su(2)) Yangian. From 
its quantum determinant we obtain an infinite series of mutually
commuting Yangian invariant operators which includes the
Hamiltonian.


\begin{center}
\end{center}


\newpage
\addtocounter{page}{-1}

\section{Introduction}

Among the exactly solvable 1d quantum systems the Hubbard model
has probably the most interesting applications in solid state
physics. Its elementary excitations, their dispersion relations
and $S$-matrix at half filling~\cite{EKS} have been calculated
exactly by use of the coordinate Bethe Ansatz in
conjunction with the SO(4) symmetry of the model~\cite{HeLi71,Ya89}. 
The story of this development is long and originates in the 
seminal
paper~\cite{LiWu68} of Lieb and Wu. A recent overview is offered by the 
reprint volume~\cite{Hubbardbook}.

\hspace{0em}From the point of view of coordinate Bethe Ansatz the 
one-dimensional Hubbard
model appears similar to the fermionic nonlinear Schr\"odinger model. 
In fact,
Lieb and Wu in their article used this analogy to obtain the
Bethe Ansatz equations from Yang's earlier result~\cite{Ya67,Ga67}.
Algebraically, however, the model seems to be more complicated.
Nearly 20 years passed before the basic tools of quantum inverse 
scattering method (QISM), $R$-matrix and $L$-matrix were derived by 
Shastry~\cite{Sha86,Sha88} and by Olmedilla et 
al.~\cite{WaOlAk87,OlWaAk87,OlWa88}, and it was shown only recently that 
the $R$-matrix satisfies the Yang-Baxter equation (YBE)~\cite{ShiWa95}.
The $R$-matrix and monodromy matrix of the
Hubbard model have unusual features. The monodromy matrix is $4
\times 4$ rather than $3 \times 3$, as one might have guessed
naively from the fact that there are two levels of Bethe Ansatz
equations or from the analogy with the fermionic nonlinear Schr\"odinger 
model.
It seems to be impossible to find a parametrization of the
$R$-matrix, such that it becomes a function of the difference
of the spectral parameters. For these reasons an algebraic Bethe
Ansatz is difficult and was performed only recently by Ramos and
Martins~\cite{RaMa96}.

There is another algebraic structure related to the Hubbard
model on the infinite line.
As was discovered by
Uglov and Korepin~\cite{UgKo94} the Hubbard Hamiltonian commutes with
two independent and mutually commuting representations of the
Yangian Y(su(2)). Yangians are the quantum groups connected to
rational solutions of the YBE~\cite{Dri86,ChaPre,LeSmi92}. The
Yangian invariance of the Hubbard Hamiltonian became likely after
the observation that the $S$-matrix of elementary excitations at
half filling is essentially a direct sum of two rational
solutions of the YBE, each corresponding to the
XXX spin chain~\cite{EsKo94}.

There is some hope that the Yangian symmetry might be used to
obtain excitation spectrum and $n$-point correlators of the
Hubbard model in a way similar to the calculation of these
quantities for the XXZ-chain by usage of its $\mbox{U}_q
(\widehat{sl_2})$ symmetry~\cite{JMMN}. Such kind of approach might
also be applicable to an extended class of non-nearest-neighbour
Hubbard models~\cite{BaGe95}, which have recently been shown to be
Yangian symmetric, too~\cite{GoIn96}, and for which a QISM approach is
unlikely to exist.

The main concern of this Letter is to show how QISM and Yangian
symmetry of the Hubbard model are connected. We benefit from the
experience of one of the authors with the fermionic nonlinear 
Schr\"odinger model~\cite{MuWa96a,MuWa96b}. 
It turns out that the situation in case of the
Hubbard model is to a large extent analogous. The Yangian
symmetry reveals, when the model is considered on the infinite
interval. Below $R$-matrix and monodromy matrix are obtained as
limits from their known counterparts on the finite interval.
The $R$-matrix greatly simplifies in the considered
limit. The new $R$-matrix contains a submatrix which turns into
the rational $R$-matrix of the XXX-chain by an appropriate
reparametrization. The corresponding submatrix of the monodromy
matrix thus provides a representation of the Y(su(2)) Yangian.
This representation is identified as the Yangian representation
constructed earlier by Uglov and Korepin using ad hoc methods. From 
the quantum determinant of the considered submatrix of the
monodromy matrix we obtain an infinite series of mutually
commuting Yangian invariant operators which is including the
Hamiltonian.

\section{Quantum Inverse Scattering Method on the Infinite Interval}
\setcounter{equation}{0}

The Hamiltonian of the one-dimensional Hubbard model is 
\begin{equation}
\hat{H}=-\sum_{j,\sigma=\uparrow,\downarrow}
(c_{j+1,\sigma}^{\dagger}c_{j,\sigma}+
c_{j,\sigma}^{\dagger}c_{j+1,\sigma})+
U\sum_{j}\left[
\left( n_{j\uparrow}-\frac{1}{2}\right)
\left( n_{j\downarrow}-\frac{1}{2}\right)
-\frac{1}{4}\right],
\label{eqn:Hamiltonian}
\end{equation}
where $c_{j,\sigma}$ and $c_{j,\sigma}^{\dagger}$ are annihilation 
and creation operators of electrons of spin $\sigma$ 
at site $j$ of a 1d lattice, and 
$n_{j\sigma}=c_{j\sigma}^{\dagger}c_{j\sigma}$ is the particle 
number operator.
Since we want to study finite excitations over the zero density vacuum 
$|0\rangle$ of the infinite interval, we normalized the Hamiltonian 
such that $\hat{H}|0\rangle=0$.

Starting point for the QISM for the Hubbard model is the exchange 
relation~\cite{OlWaAk87}
\begin{equation}
{\cal R}(\lambda,\mu)[ {\cal L}_{m}(\lambda)
\otimes_{\mathrm{s}} {\cal L}_{m}(\mu) ]
=[ {\cal L}_{m}(\mu)\otimes_{\mathrm{s}} {\cal L}_{m}(\lambda) ]
{\cal R}(\lambda,\mu),
\label{eqn:RLL}
\end{equation}
where 
$\otimes_{\mathrm{s}}$ denotes the Grassmann direct product
\begin{equation}
[A\otimes_{\mathrm{s}} B]_{\alpha\gamma,\beta\delta}=
(-1)^{[P(\alpha)+P(\beta)]P(\gamma)} A_{\alpha\beta}B_{\gamma\delta}
\end{equation}
with grading $P(1)=P(4)=0,\; P(2)=P(3)=1$.
We adopt the expressions for the matrices ${\cal R}$ and 
${\cal L}$ in terms of 
two parametrizing functions $\alpha(\lambda), \;\gamma(\lambda)$ 
from ref.~\cite{OlWaAk87}. For later convenience, however, we 
shift the arguments of $\alpha(\lambda)$ and $\gamma(\lambda)$ 
by $\frac{\pi}{4}$, such that we simply have $\alpha(\lambda)=\cos \lambda$, 
$\gamma(\lambda)=\sin \lambda $. The $L$-matrix is 
\begin{equation}
{\cal L}_{m}(\lambda)=
\left(
\begin{array}{cccc}
-\mathrm{e}^{h(\lambda)}f_{m\uparrow}f_{m\downarrow} & 
-f_{m\uparrow}c_{m\downarrow} & \mathrm{i}c_{m\uparrow}f_{m\downarrow} 
&\mathrm{i}c_{m\uparrow}c_{m\downarrow} \mathrm{e}^{h(\lambda)} 
\\
-\mathrm{i}f_{m\uparrow}c_{m\downarrow}^{\dagger} 
&  \mathrm{e}^{-h(\lambda)}f_{m\uparrow}g_{m\downarrow} 
&  \mathrm{e}^{-h(\lambda)}c_{m\uparrow}c_{m\downarrow}^{\dagger} 
& \mathrm{i}c_{m\uparrow}g_{m\downarrow} 
\\
c_{m\uparrow}^{\dagger}f_{m\downarrow} 
& \mathrm{e}^{-h(\lambda)}c_{m\uparrow}^{\dagger}c_{m\downarrow}
& \mathrm{e}^{-h(\lambda)} g_{m\uparrow}f_{m\downarrow} 
& g_{m\uparrow}c_{m\downarrow} 
\\
-\mathrm{i}\mathrm{e}^{h(\lambda)}c_{m\uparrow}^{\dagger}
c_{m\downarrow}^{\dagger}
& c_{m\uparrow}^{\dagger}g_{m\downarrow}
& \mathrm{i}g_{m\uparrow}c_{m\downarrow}^{\dagger}
&-g_{m\uparrow}g_{m\downarrow} \mathrm{e}^{h(\lambda)}
\end{array}
\right) ,
\end{equation}
where 
$f_{m\sigma}(\lambda)=\gamma(\lambda)(1-n_{m\sigma})+
\mathrm{i}\alpha(\lambda)n_{m\sigma}$, 
$g_{m\sigma}(\lambda)=\alpha(\lambda)(1-n_{m\sigma})-
\mathrm{i}\gamma(\lambda)n_{m\sigma}$, and  
$h(\lambda)$ is defined as 
\begin{equation}
\frac{\sinh 2h(\lambda)}{\sin 2\lambda}=\frac{U}{4}.
\label{eqn:U4}
\end{equation}
Due to space limitations we do not reproduce the $R$-matrix here. 
It is $16\times 16$ and contains 36 nonvanishing 
entries, only ten of which are different modulo signs.
The ten different entries are denoted by $\rho_{i}$, $i=1,\cdots,10$, 
in ref.~\cite{OlWaAk87}. 
They are rational functions of $\alpha(\lambda)$, $\gamma(\lambda)$ 
and $\mathrm{e}^{h(\lambda)}$. We provide a list and some basic 
formulae which have been used in our calculations in \ref{appendix:A}.
The $(m-n)$-site Hubbard model $(m>n)$ is characterized by 
the monodromy matrix 
\begin{equation}
{\cal T}_{mn}(\lambda)={\cal L}_{m-1}(\lambda){\cal L}_{m-2}
(\lambda)\cdots {\cal L}_{n}(\lambda).
\end{equation}
It has been shown in ref.~\cite{Sha86,OlWaAk87} 
that the logarithmic derivative of the 
graded trace of ${\cal T}_{mn}(\lambda)$ at $\lambda =0$ 
reproduces the Hamiltonian (\ref{eqn:Hamiltonian}) 
under periodic boundary conditions. Like ${\cal L}_{m}(\lambda)$  
the monodromy matrix ${\cal T}_{mn}(\lambda)$ satisfies (\ref{eqn:RLL}).
In contrast to the classical case~\cite{FaTa} a formulation 
of QISM on the infinite interval~\cite{Fa80,Tha81} has so far 
only been possible for zero density of elementary particles. 
This is due to the complicated structure of the finite density 
(finite band filling) vacua formed by an infinite number of 
interacting particles. For the Hubbard model, 
there are four simple vacua, 
the empty band, the completely filled band and the half filled band 
with all spins up or all spins down. In the following we will consider 
the empty band $|0\rangle$ (
$c_{m\sigma}|0\rangle=0$) 
as reference state. 
Zero particle density means to 
consider only states with a finite number of particles in the empty 
band. Expectation values of the $L$-matrix with respect to 
this space have a finite limit for $|m|\rightarrow\infty$, 
which formally
can be obtained by setting normal ordered products of 
operators equal to zero~\cite{Fa80,Tha81}, 
\begin{equation}
{\cal L}(\lambda)\rightarrow V(\lambda)=\mathrm{diag} 
(-\gamma(\lambda)^{2}\mathrm{e}^{h(\lambda)},
\alpha(\lambda)\gamma(\lambda)\mathrm{e}^{-h(\lambda)},
\alpha(\lambda)\gamma(\lambda)\mathrm{e}^{-h(\lambda)},
-\alpha(\lambda)^{2}\mathrm{e}^{h(\lambda)}).
\end{equation}
$V(\lambda)$ can be used to split off the asymptotics of 
${\cal T}_{mn}(\lambda)$. We expect the matrix 
\begin{equation}
\tilde{{\cal T}}_{mn}(\lambda)=
V(\lambda)^{-m}{\cal T}_{mn}(\lambda)
V(\lambda)^{n}
\label{eqn:tildeT}
\end{equation}
to have a finite limit for $m\rightarrow\infty$, $n\rightarrow -\infty$.
This limit, 
\begin{equation}
\tilde{{\cal T}}(\lambda)=
\lim_{m,-n\rightarrow\infty} \tilde{{\cal T}}_{mn}(\lambda),
\label{eqn:deftildeT}
\end{equation}
will be the monodromy matrix on the infinite interval. 
To derive an exchange relation for $\tilde{{\cal T}}(\lambda)$, 
consider the asymptotics $W(\lambda,\mu)$ of 
${\cal L}_{m}(\lambda)\otimes_{\mathrm{s}} 
{\cal L}_{m}(\mu)$ for large $|m|$ 
again by omitting all normal ordered products of Fermi operators. 
Then $W(\lambda,\mu)^{-m}({\cal T}_{mn}(\lambda)\otimes_{\mathrm{s}}
{\cal T}_{mn}(\mu))W(\lambda,\mu)^{n}$ is expected to have a 
finite limit. $W(\lambda,\mu)$ is not just the tensor product 
$V(\lambda)\otimes_{\mathrm{s}}V(\mu)$.
Due to normal ordering there appear additional off-diagonal 
elements, 
\begin{eqnarray*}
\lefteqn{W(\lambda,\mu)_{12,21}=W(\lambda,\mu)_{13,31}=
-\mathrm{i}\gamma(\lambda)\gamma(\mu), \makebox[1em]{}
W(\lambda,\mu)_{14,23}=-W(\lambda,\mu)_{14,32}=
-\mathrm{i}\gamma(\lambda)\alpha(\mu),} \\
&& \hspace{-2em} W(\lambda,\mu)_{24,42}=W(\lambda,\mu)_{34,43}=
\mathrm{i}\alpha(\lambda)\alpha(\mu), \makebox[1em]{}
W(\lambda,\mu)_{23,41}=-W(\lambda,\mu)_{32,41}=
-\mathrm{i}\alpha(\lambda)\gamma(\mu), \\
&& \hspace{-2em}
W(\lambda,\mu)_{14,41}=-\mathrm{e}^{h(\lambda)+h(\mu)}.
\end{eqnarray*}
It follows from the exchange relation for the monodromy 
matrix that 
\begin{equation}
{\cal R}(\lambda,\mu)W(\lambda,\mu)=W(\mu,\lambda){\cal R}(\lambda,\mu).
\label{eqn:RW}
\end{equation}
Using once more the exchange relation and 
the definition (\ref{eqn:tildeT})
of $\tilde{{\cal T}}_{mn}(\lambda)$ we obtain 
\begin{eqnarray}
&&U_{m}(\mu,\lambda)^{-1}{\cal R}(\lambda,\mu)U_{m}(\lambda,\mu)
[\tilde{{\cal T}}_{mn}(\lambda)\otimes_{\mathrm{s}}
\tilde{{\cal T}}_{mn}(\mu)]
\nonumber \\
&& \makebox[1em]{} 
=[\tilde{{\cal T}}_{mn}(\mu)\otimes_{\mathrm{s}}
\tilde{{\cal T}}_{mn}(\lambda)]
U_{n}(\mu,\lambda)^{-1}{\cal R}(\lambda,\mu)
U_{n}(\lambda,\mu),
\label{eqn:URU}
\end{eqnarray}
where 
\begin{equation}
U_{m}(\lambda,\mu)=W(\lambda,\mu)^{-m}[V(\lambda)^{m} 
\otimes_{\mathrm{s}}V(\mu)^{m}].
\end{equation}
Postponing the discussion of convergence for while we
formally take the limits $m,-n\rightarrow\infty$ in (\ref{eqn:URU}).
Then by use of the definitions 
\begin{equation}
U_{\pm}(\lambda,\mu)=\lim_{m\rightarrow\pm\infty}U_{m}(\lambda,\mu), 
\makebox[1em]{} 
\tilde{\cal R}^{(\pm)}(\lambda,\mu)=
U_{\pm}(\mu,\lambda)^{-1}{\cal R}(\lambda,\mu)
U_{\pm}(\lambda,\mu), 
\end{equation}
we arrive at the exchange relation for the monodromy matrix 
$\tilde{{\cal T}}(\lambda)$ on the infinite interval, 
\begin{equation}
\tilde{R}^{(+)}(\lambda,\mu)\left[
\tilde{\cal T}(\lambda)\otimes_{\mathrm{s}}\tilde{\cal T}(\mu)\right]
=\left[\tilde{\cal T}(\mu)
\otimes_{\mathrm{s}}\tilde{\cal T}(\lambda)\right]
\tilde{R}^{(-)}(\lambda,\mu).
\label{eqn:newRTT}
\end{equation}

The matrices $U_{\pm}(\lambda,\mu)$, their inverses and the
$R$-matrices $\tilde{{\cal R}}^{(\pm)}(\lambda,\mu)$ can be calculated 
as functions of the $\rho_{i}$'s by utilizing the formulae 
provided in \ref{appendix:A}. 
The non-zero matrix elements of 
$U_{\pm}(\lambda,\mu)$ follow as 
\begin{eqnarray}
&& U_{\pm}(\lambda,\mu)_{\alpha\beta,\alpha\beta}=1, \;\;
U_{\pm}(\lambda,\mu)_{14,41}=\frac{-\rho_{5}}{\rho_{5}-\rho_{4}}, 
\nonumber \\
&& U_{\pm}(\lambda,\mu)_{12,21}=
U_{\pm}(\lambda,\mu)_{13,31}=\frac{-\mathrm{i}\rho_{2}}{\rho_{10}}, \;\;
U_{\pm}(\lambda,\mu)_{14,23}=
-U_{\pm}(\lambda,\mu)_{14,32}=
\frac{\mathrm{i}\rho_{6}}{\rho_{3}-\rho_{1}}, 
\nonumber \\
&& U_{\pm}(\lambda,\mu)_{23,41}=
-U_{\pm}(\lambda,\mu)_{32,41}=
\frac{\mathrm{i}\rho_{6}}{\rho_{5}-\rho_{4}},\;\;
U_{\pm}(\lambda,\mu)_{24,42}=
U_{\pm}(\lambda,\mu)_{34,43}=\frac{\mathrm{i}\rho_{2}}{\rho_{9}},
\nonumber
\end{eqnarray}
where $\rho_{i}=\rho_{i}(\lambda,\mu)$.
The corresponding matrices $\tilde{\cal R}^{(\pm)}$ are 
\begin{eqnarray}
\lefteqn{\tilde{\cal R}^{(+)}(\lambda,\mu)= 
\tilde{\cal R}^{(-)}(\lambda,\mu) = }\nonumber \\[1ex]
& & \hspace{-22pt} \left( 
{\arraycolsep 2pt
\begin{array}{cccccccccccccccc}
\rho_{1} & 0 & 0 & 0 & 
0 & 0 & 0 & 0 & 0 & 0 & 0 & 0 & 0 & 0 & 0 & 0 \\
0 & 0 & 0 & 0 & 
\frac{\rho_{1}\rho_{4}}{\mathrm{i}\rho_{10}} 
& 0 & 0 & 0 & 0 & 0 & 0 & 0 
& 0 & 0 & 0 & 0 \\
0 & 0 & 0 & 0 & 
0 & 0 & 0 & 0 &
\frac{\rho_{1}\rho_{4}}{\mathrm{i}\rho_{10}} 
& 0 & 0 & 0 & 0 & 0 & 0 & 0 \\
0 & 0 & 0 & 0 & 
0 & 0 & 0 & 0 &
0 & 0 & 0 & 0 &
\frac{-\rho_{1}\rho_{4}}{\rho_{5}-\rho_{4}} & 0 & 0 & 0 \\
0 & 
 -\mathrm{i}\rho_{10} &0 & 0 & 0 &  0 & 0 & 0 &
0 & 0 & 0 & 0 &
0 & 0 & 0 & 0 \\
0 & 0 & 0 & 0 & 
0 &  \rho_{4} & 0 & 0 &
0 & 0 & 0 & 0 &
0 & 0 & 0 & 0 \\
0 & 0 & 0 & 0 & 
0 & 0 & \frac{\rho_{3}\rho_{4}-\rho_{2}^{2}}{\rho_{3}-\rho_{1}}
& 0 & 0 & 
\frac{\rho_{9}\rho_{10}}{\rho_{3}-\rho_{1}} & 0 & 0 &
0 & 0 & 0 & 0 \\
0 & 0 & 0 & 0 & 
0 & 0 & 0 & 0 &
0 & 0 & 0 & 0 &
0 & 
\frac{\mathrm{i}\rho_{1}\rho_{4}}{\rho_{9}}
 & 0 & 0 \\
0 & 0 &  -\mathrm{i}\rho_{10} & 0 & 
0 & 0 & 0 & 0 &
0 & 0 & 0 & 0 &
0 & 0 & 0 & 0 \\
0 & 0 & 0 & 0 & 
0 & 0 &\frac{\rho_{9}\rho_{10}}{\rho_{3}-\rho_{1}}
& 0 & 0 &  \frac{\rho_{3}\rho_{4}-\rho_{2}^{2}}{\rho_{3}-\rho_{1}}
 & 0 & 0 & 
0 & 0 & 0 & 0 \\
0 & 0 & 0 & 0 & 
0 & 0 & 0 & 0 &
0 & 0 &  \rho_{4} & 0 &
0 & 0 & 0 & 0 \\
0 & 0 & 0 & 0 & 
0 & 0 & 0 & 0 &
0 & 0 & 0 & 0 &
0 & 0 & \frac{\mathrm{i}\rho_{1}\rho_{4}}{\rho_{9}}
& 0 \\
0 & 0 & 0 &  \rho_{1}-\rho_{3} & 
0 & 0 & 0 & 0 &
0 & 0 & 0 & 0 &
0 & 0 & 0 & 0 \\
0 & 0 & 0 & 0 & 
0 & 0 & 0 &  \mathrm{i} \rho_{9} &
0 & 0 & 0 & 0 &
0 & 0 & 0 & 0 \\
0 & 0 & 0 & 0 & 
0 & 0 & 0 & 0 &
0 & 0 & 0 &  \mathrm{i} \rho_{9} &
0 & 0 & 0 & 0 \\
0 & 0 & 0 & 0 & 
0 & 0 & 0 & 0 &
0 & 0 & 0 & 0 &
0 & 0 & 0 &  \rho_{1} 
\end{array}
}
\right)
\raisebox{-22ex}{.} \makebox[2em]{}
\label{eqn:deftildeR}
\end{eqnarray}
The reader is urged to compare this expression with the $R$-matrix on the 
finite interval~\cite{OlWaAk87}. Instead of the 
36 nonvanishing elements of 
the original $R$-matrix we have only 18 nonvanishing elements here, 
which brings about simpler commutation relations between
elements of the monodromy matrix.

\hspace{0em}From our experience with the finite interval case we know 
that the monodromy matrix is of the following block form, 
\begin{equation}
\tilde{{\cal T}}(\lambda)=
\left(
\begin{array}{cccc}
D_{11}(\lambda) & C_{11}(\lambda) & C_{12}(\lambda) & D_{12}(\lambda) \\
B_{11}(\lambda) & A_{11}(\lambda) &A_{12}(\lambda) & B_{12}(\lambda) \\
B_{21}(\lambda) & A_{21}(\lambda) & A_{22}(\lambda) & B_{22}(\lambda) \\
D_{21}(\lambda) & C_{21}(\lambda) & C_{22}(\lambda) & D_{22}(\lambda) 
\end{array} 
\right),
\label{eqn:tildeTelements}
\end{equation}
where on the finite interval $A(\lambda)$ corresponds to the 
su(2) Lie algebra of rotations, $D(\lambda)$ corresponds to the 
$\eta$-pairing su(2) Lie algebra and the blocks $B(\lambda)$ 
and $C(\lambda)$ are connected to each other by particle-hole and 
gauge transformations~\cite{GoMu96}. Using the explicit form of the 
matrices $\tilde{{\cal R}}^{(\pm)}$ the exchange relation 
(\ref{eqn:tildeTelements}) implies
\begin{equation}
\frac{\rho_{9}(\lambda,\mu)\rho_{10}(\lambda,\mu)}{
\rho_{3}(\lambda,\mu)\rho_{4}(\lambda,\mu)-\rho_{2}(\lambda,\mu)^{2}}
\left[
A_{\alpha\beta}(\lambda),A_{\gamma\delta}(\mu) \right]
=A_{\gamma\beta}(\mu)A_{\alpha\delta}(\lambda)
-A_{\gamma\beta}(\lambda)A_{\alpha\delta}(\mu),
\label{eqn:AAAA}
\end{equation}
i.e.\ the commutation relations between the matrix elements of 
$A(\lambda)$  are decoupled from the rest of the algebra. This fact 
will be crucial for the derivation of the Y(su(2)) Yangian 
representation below. 

So far we avoided to comment on the analytic structure of the exchange 
relation (\ref{eqn:newRTT}). This is indeed a delicate point. 
Along with the calculation of the limits $U_{\pm}(\lambda,\mu)$ we 
obtain convergence conditions. The convergence conditions for 
$U_{+}(\lambda,\mu)$ and $U_{-}(\lambda,\mu)$ are complementary. 
This situation is typical for the QISM on the infinite interval.
Thus the first equation in (\ref{eqn:deftildeR}) is only formal 
for the present time. 
We conjecture, however, that analytic continuation in $\lambda$ and
$\mu$ respects the exchange relation (\ref{eqn:newRTT}) with the 
possible exception of $\lambda=\mu$ (modulo periods), where 
singular terms (like $\delta(\lambda-\mu)$ ) may destroy the 
first equality in (\ref{eqn:deftildeR}). We know from our experience 
with the fermionic nonlinear Schr\"odinger model~\cite{MuWa96a,MuWa96b} 
that such singular terms are irrelevant for the derivation of 
a Yangian representation from the exchange relation 
(\ref{eqn:newRTT}).

\section{Yangian Symmetry}
\setcounter{equation}{0}

The Y(su(2)) Yangian~\cite{Dri86,ChaPre,LeSmi92}
algebra is generated by six generators $Q_{n}^{a} \;(n=0,1;a=1,2,3)$, 
satisfying the following relations;
\begin{eqnarray}
& & \left[ Q_{0}^{a}, Q_{0}^{b} \right]  =  f^{abc}Q_{0}^{c}, 
\label{eqn:Y1} \\
& & \left[ Q_{0}^{a}, Q_{1}^{b} \right]  =  f^{abc}Q_{1}^{c}, 
\label{eqn:Y2} \\
& & \left[ \left[ Q_{1}^{a},Q_{1}^{b} \right], 
\left[ Q_{0}^{c},Q_{1}^{d} \right] \right]+
\left[ \left[ Q_{1}^{c},Q_{1}^{d} \right], 
\left[ Q_{0}^{a},Q_{1}^{b} \right] \right] \nonumber \\
& & \makebox[2em]{}
=\kappa^{2}(A^{abkefg}f^{cdk}+A^{cdkefg}f^{abk})
\{Q_{0}^{e}, Q_{0}^{f}, Q_{1}^{g} \},
\label{eqn:Y3}
\end{eqnarray}
where $\kappa$ is a nonzero constant, 
$\sigma^{a}\; (a=1,2,3)$ are the Pauli matrices, 
$f^{abc}=i\varepsilon^{abc}$ is the antisymmetric tensor of 
structure constants 
of su(2), and 
$A^{abcdef}=f^{adk}f^{bel}f^{cfm}f^{klm}$.
Here and in the following 
we are using implicit summation over 
doubly occuring indices. The bracket $\{\; \; \}$ denotes the 
symmetrized product 
\begin{equation}
\{ x_{1}, \cdots, x_{m}\} =\frac{1}{m!} \sum_{\sigma\in S_{m}}
x_{\sigma 1}\cdots x_{\sigma m}.
\end{equation}

Now we will show how to obtain a representation of Y(su(2)) 
from the submatrix $A(\lambda)$ of the monodromy matrix 
(\ref{eqn:tildeTelements}). 
Introducing the reparametrization 
\begin{equation}
v(\lambda)=-2\mathrm{i} \, \cot 2 \lambda\cosh 2h(\lambda),
\label{eqn:defv}
\end{equation}
the prefactor on the lhs of (\ref{eqn:AAAA}) becomes 
\begin{equation}
\frac{\rho_{9}(\lambda,\mu)\rho_{10}(\lambda,\mu)}{
\rho_{3}(\lambda,\mu)\rho_{4}(\lambda,\mu)-\rho_{2}(\lambda,\mu)^{2}}
=\frac{ v(\lambda)-v(\mu) }{\mathrm{i}U},
\label{eqn:rho910}
\end{equation}
and we can write (\ref{eqn:AAAA}) in matrix form as 
\begin{equation}
\left( \mathrm{i}U+\{ v(\lambda)-v(\mu)\} {\cal P}\right)
[A(\lambda)\otimes A(\mu)]
=[A(\mu)\otimes A(\lambda)]
\left( \mathrm{i}U+\{ v(\lambda)-v(\mu)\} {\cal P}\right).
\label{eqn:commutA}
\end{equation}
Here ${\cal P}$ is a $4\times 4$ permutation matrix.
$(\mathrm{i}U+\{ v(\lambda)-v(\mu)\} {\cal P})$ is the 
rational $R$-matrix of the XXX spin chain. 
(\ref{eqn:commutA}) implies that 
$A_{\alpha\beta}(\lambda)$ is a generating function of the 
Y(su(2)) Yangian.
The Yangian generators $Q_{0}^{a}$ and $Q_{1}^{a}$ are the first 
coefficients in the asymptotic 
expansion~\cite{Hal94,MuWa96b}
\begin{equation}
A_{\alpha\beta}(\lambda)=1+\mathrm{i}U \sum_{n=0}^{\infty}
\frac{1}{v(\lambda)^{n+1}}
\left( \sum_{a=1}^{3}
Q_{n}^{a}\tilde{\sigma}_{\alpha\beta}^{a}+Q_{n}^{0}\delta_{\alpha\beta}
\right).
\label{eqn:expandA}
\end{equation}
In order to obtain compact expression for $Q_{0}^{a}$ and $Q_{1}^{a}$ 
we introduced the abbreviations
$\tilde{\sigma}^{1}=-\sigma^{2}$, 
$\tilde{\sigma}^{2}=\sigma^{1}$, 
$\tilde{\sigma}^{3}=\sigma^{3}$.\footnote{
The reader should not worry about this notation. It results from our 
choice of the $L$-matrix, which we took from ref.~\cite{OlWaAk87} 
to facilitate comparison with earlier work.
It is easy to introduce a slight change of the $L$-matrix, compatible
with the exchange relation, such that $\tilde{\sigma}^{a}$ is 
replaced by $\sigma^{a}$ in (\ref{eqn:expandA}).
}
There are several possibilities to perform the limit 
$v(\lambda)\rightarrow\infty$. However, we found that only 
one of these yields finite results for $Q_{0}^{a}$ and $Q_{1}^{a}$. 
We have to take $\mathrm{Im}(\lambda)\rightarrow\infty$ and 
further have to choose the proper branch of solution in 
eqn.(\ref{eqn:U4}) which determines $h(\lambda)$ as a function of 
$\lambda$. Some of the details of the calculation are given in 
\ref{appendix:B}. The final result is 
\begin{eqnarray}
Q_{0}^{a}&=& \frac{1}{2}\sum_{j}\sigma_{\alpha\beta}^{a}
a^{\dagger}_{j,\alpha}c_{j,\beta}, 
\label{eqn:Q0}
\\
Q_{1}^{a}&=& -\frac{\mathrm{i}}{2}\sum_{j}\sigma_{\alpha\beta}^{a}
c_{j,\alpha}^{\dagger}(c_{j+1,\beta}-c_{j-1,\beta})
-\frac{\mathrm{i}U}{4}\sum_{i,j}
\mathrm{sgn}(j-i)\sigma_{\alpha\beta}^{a}
c_{i,\alpha}^{\dagger}c_{j,\gamma}^{\dagger}
c_{i,\gamma}c_{j,\beta}.
\label{eqn:Q1}
\end{eqnarray}
In our conventions 
the constant $\kappa$ in (\ref{eqn:Y3}) is equal to $\mathrm{i}U$.
Comparing (\ref{eqn:Q0}), (\ref{eqn:Q1}) with the result of 
Uglov and Korepin~\cite{UgKo94}, we find complete equivalence.  
\begin{eqnarray}
&&E_{0}=Q_{0}^{1}+\mathrm{i}Q_{0}^{2}, \;
F_{0}=Q_{0}^{1}-\mathrm{i}Q_{0}^{2}, \;
H_{0}=2Q_{0}^{3}, \\
&&E_{1}=Q_{1}^{1}+\mathrm{i}Q_{1}^{2}, \;
F_{1}=Q_{1}^{1}-\mathrm{i}Q_{1}^{2}, \;
H_{1}=2Q_{1}^{3}, 
\end{eqnarray}
where the expressions on the lhs are taken from the 
paper of Uglov and Korepin.

Associated with the exchange relation (\ref{eqn:commutA})
we can consider the quantum determinant~\cite{IzKo81,KBIbook}, 
\begin{equation}
\mathrm{Det}_{q}A(\lambda)=
A_{11}(v(\lambda))A_{22}\left( 
v(\lambda)-\mathrm{i}U \right)
-A_{12}(v(\lambda))A_{21}\left(
v(\lambda)-\mathrm{i}U \right),
\end{equation}
which is in the center of the Yangian and provides a generating 
function of mutually commuting operators, 
\begin{equation}
[\mathrm{Det}_{q}A(\lambda), A_{\alpha\beta}(\mu)]=0, 
\makebox[1em]{}
[\mathrm{Det}_{q}A(\lambda), \mathrm{Det}_{q}A(\mu)]=0. 
\label{eqn:DetAA}
\end{equation}
The second of these equations is of course a consequence of 
the first one. Performing again the asymptotic 
expression in terms of $v(\lambda)$, 
\begin{equation}
\mathrm{Det}_{q}A(\lambda)= 1+\mathrm{i}U\sum_{n=0}^{\infty}
\frac{1}{v(\lambda)^{n+1}}I_{n}, 
\end{equation}
we obtain 
$I_{0}=0$, $I_{1}=\mathrm{i}\hat{H}$, 
i.e.\ the Hamiltonian is among the commuting operators.
All the conserved operators are Yangian invariant by construction. 
It will be interesting to investigate their relation to
the formerly known conserved quantities~\cite{Sha86,Sha88,OlWa88,Gro},  
which were obtained for the finite periodic model.

In closing this section we shall add a comment. 
The Hubbard Hamiltonian on the infinite interval is invariant under 
the transformation
\begin{equation}
c_{j,\downarrow}\rightarrow c_{j,\downarrow}, \; 
c_{j,\uparrow}\rightarrow (-1)^{j}c_{j,\uparrow}^{\dagger}, \;
U\rightarrow -U.
\end{equation}
The Yangian generators $Q_{0}^{a}$ and $Q_{1}^{a}$, 
however, are transformed into a pair of generators
$Q_{0}^{a\prime}$ and $Q_{1}^{a\prime}$ of a second, independent 
representation of Y(su(2))~\cite{UgKo94}. These two representations 
mutually commute. Therefore they can be combined to a 
direct sum Y(su(2))$\oplus$Y(su(2)). The reason why we get only 
one of these representations from our QISM approach is that, 
in order to perform the passage to the infinite interval, we 
refer to the zero density vacuum $|0\rangle$. This vacuum has lower 
symmetry than the Hamiltonian. It is invariant under the su(2) Lie 
algebra of rotations, but does not respect the $\eta$-pairing 
su(2) symmetry of the Hamiltonian. 
A fully su(2)$\oplus$su(2) invariant vacuum would be the singlet
ground state at half filling~\cite{EsKo94}.
It seems to be yet a formidable task to formulate the 
QISM with respect to this state.

\section{Concluding Remarks and Discussion}
\setcounter{equation}{0}

We have developed the QISM for the Hubbard model on 
the infinite interval with respect to the zero density vacuum. 
The $R$-matrix (\ref{eqn:deftildeR}) thus obtained
is greatly simplified in comparison with the $R$-matrix 
of the finite periodic model. 
Particularly, it reveals a hidden rational structure, 
which arises from a certain combination (\ref{eqn:rho910}) of 
the functions $\rho_{i}$. This structure was discovered earlier 
by Ramos and Martins~\cite{RaMa96} as part of the exchange relation 
for the Hubbard model on the finite interval. 
Note, however, that our reparametrization (\ref{eqn:rho910}), 
which is essentially unique, differs from that given in \cite{RaMa96}. 
A comparison is obstructed by the fact that the authors do not 
expose their parameters $\alpha_{j}$. Along with the simplified 
$R$-matrix we obtained the 
aymptotic expansion (\ref{eqn:expandA}) of the 
submatrix $A(\lambda)$ of the 
monodromy matrix, which naturally provides a representation 
of Y(su(2)) 
and generates an infinite series of mutually commuting 
Yangian invariant operators including the Hamiltonian.

There is a number of interesting open problems related to the QISM 
on the infinite interval. The analytic properties of the $R$-matrices 
$\tilde{\cal R}^{(\pm)}$ (\ref{eqn:deftildeR}) and of the monodromy 
matrix (\ref{eqn:deftildeT}) deserve further investigations. 
Only the submatrix $A(\lambda)$ of the monodromy matrix has a limit 
for ${\mathrm Im}(\lambda)\rightarrow\infty$. All other matrix elements 
diverge. It is therefore not clear at the present stage of investigation 
how to obtain creation operators of elementary excitations that 
are compatible with the Yangian generators $Q_{0}^{a}$ and $Q_{1}^{a}$. 
Creation operators are indispensable for the discussion of irreducible 
Yangian representations~\cite{MuWa96a,MuWa96b}. 

Another interesting task will be the construction of Dunkl 
operators~\cite{Dunkl} associated with the Yangian representation 
discussed in this Letter. 
Dunkl operators are building blocks of 
Yangian generators~\cite{Dunkl,HiMu96,MuWa96a,MuWa96b}. 
They are useful for the investigation of eigenstates. For the 
one-dimensional Hubbard model, although several 
attempts~\cite{Ug95,HiMu96} have been made, 
no satisfactory Dunkl operator is known.

\section*{Acknowledgements}
We are grateful to Professor Miki Wadati for continuous encouragement 
and comments. 
We would like to thank 
K.~Hikami, V.~E.~Korepin and M.~Shiroishi for fruitful discussions and 
comments. 
S.~M. is also grateful to Professor Naoto Nagaosa for his encouragement. 
This work has been supported by the Japan Society for the 
Promotion of Science and the Ministry of Science, Culture and Education of 
Japan.

\setcounter{section}{0}
\renewcommand{\thesection}{Appendix \Alph{section}}
\section{Relations Between the Elements of the 
$R$-Matrix}
\label{appendix:A}
\renewcommand{\theequation}{\Alph{section}.\arabic{equation}}
\setcounter{equation}{0}

In this appendix 
we collect functional relations among the elements of 
the $R$-matrix, which have been used in the calculation 
of the matrix $U_{\pm}(\lambda,\mu)$. 
We begin with the defining relations of the 
matrix elements, which are
\begin{eqnarray}
\frac{\rho_{1}(\lambda, \mu)}{\rho_{2}(\lambda, \mu)} &=&
\mathrm{e}^{l}\alpha(\lambda)\alpha(\mu)+
\mathrm{e}^{-l}\gamma(\lambda)\gamma(\mu), \label{eqn:rho1} \\
\frac{\rho_{4}(\lambda, \mu)}{\rho_{2}(\lambda, \mu)} &=&
\mathrm{e}^{l}\gamma(\lambda)\gamma(\mu)+
\mathrm{e}^{-l}\alpha(\lambda)\alpha(\mu), \label{eqn:rho2} \\
\frac{\rho_{9}(\lambda, \mu)}{\rho_{2}(\lambda, \mu)} &=&
-\mathrm{e}^{l}\alpha(\lambda)\gamma(\mu)+
\mathrm{e}^{-l}\gamma(\lambda)\alpha(\mu), \label{eqn:rho9} \\
\frac{\rho_{10}(\lambda, \mu)}{\rho_{2}(\lambda, \mu)} &=&
\mathrm{e}^{l}\gamma(\lambda)\alpha(\mu)-
\mathrm{e}^{-l}\alpha(\lambda)\gamma(\mu), \label{eqn:rho10} \\
\frac{\rho_{3}(\lambda, \mu)}{\rho_{2}(\lambda, \mu)} &=&
\frac{\mathrm{e}^{l}\alpha(\lambda)\alpha(\mu)-
\mathrm{e}^{-l}\gamma(\lambda)
\gamma(\mu)}{\alpha^{2}(\lambda)-\gamma^{2}(\mu)},
\label{eqn:rho3} \\
\frac{\rho_{5}(\lambda, \mu)}{\rho_{2}(\lambda, \mu)} &=&
\frac{-\mathrm{e}^{l}\gamma(\lambda)\gamma(\mu)+
\mathrm{e}^{-l}\alpha(\lambda)
\alpha(\mu)}{\alpha^{2}(\lambda)-\gamma^{2}(\mu)},
\label{eqn:rho5} \\
\frac{\rho_{6}(\lambda, \mu)}{\rho_{2}(\lambda, \mu)} &=&
\frac{\mathrm{e}^{-h} [ \mathrm{e}^{l}\alpha(\lambda)\gamma(\lambda)-
\mathrm{e}^{-l}\alpha(\mu)
\gamma(\mu)] }{\alpha^{2}(\lambda)-\gamma^{2}(\mu)},
\label{eqn:rho6}
\end{eqnarray}
where $h=h(\lambda)+h(\mu), \;l=h(\lambda)-h(\mu)$.
There are relations among the $\rho_{i}$ functions; 
\begin{eqnarray}
\rho_{1}(\lambda,\mu)\rho_{4}(\lambda,\mu)+
\rho_{9}(\lambda,\mu)\rho_{10}(\lambda,\mu)
&=& \rho_{2}(\lambda,\mu)^{2}, \label{eqn:rhorel1} \\
\rho_{3}(\lambda,\mu)\rho_{5}(\lambda,\mu)-
\rho_{6}(\lambda,\mu)^{2}
&=& \rho_{2}(\lambda,\mu)^{2}, \label{eqn:rhorel2}\\
\rho_{1}(\lambda,\mu)\rho_{5}(\lambda,\mu)+
\rho_{3}(\lambda,\mu)\rho_{4}(\lambda,\mu)
&=& 2\rho_{2}(\lambda,\mu)^{2}. \label{eqn:rhorel3}
\end{eqnarray}
We found that there is a set of relations "dual" to 
(\ref{eqn:rho1})-(\ref{eqn:rho6}).
Introducing a transformation $\Phi$, which 
keeps $\lambda$ unchanged and substitutes $\mu+\frac{\pi}{2}$ for 
$\mu$, 
we get the following transformation rules; 
\begin{eqnarray*}
& & \frac{\rho_{1}}{\rho_{2}} \rightarrow 
\frac{\rho_{5}-\rho_{4}}{\rho_{6}}, \makebox[2em]{} 
\frac{\rho_{4}}{\rho_{2}} \rightarrow 
\frac{\rho_{3}-\rho_{1}}{\rho_{6}}, \makebox[2em]{}
\frac{\rho_{9}}{\rho_{2}} \rightarrow 
-\frac{\rho_{10}}{\rho_{6}}, \makebox[2em]{} 
\frac{\rho_{10}}{\rho_{2}} \rightarrow 
-\frac{\rho_{9}}{\rho_{6}}, \\
& & \frac{\rho_{3}}{\rho_{2}} \rightarrow 
\frac{\rho_{5}}{\rho_{6}}, \makebox[4em]{} 
\frac{\rho_{5}}{\rho_{2}} \rightarrow 
\frac{\rho_{3}}{\rho_{6}}, \makebox[4em]{}
\frac{\rho_{6}}{\rho_{2}} \rightarrow 
-\frac{\rho_{2}}{\rho_{6}}. 
\end{eqnarray*}
Explicitly, we obtain the relations
\begin{eqnarray}
\frac{\rho_{5}(\lambda, \mu)-\rho_{4}(\lambda, \mu)}{\rho_{6}(\lambda, \mu)} &=&
-\mathrm{e}^{h}\alpha(\lambda)\gamma(\mu)+
\mathrm{e}^{-h}\gamma(\lambda)\alpha(\mu),  \label{eqn:rho1'} \\
\frac{\rho_{3}(\lambda, \mu)
-\rho_{1}(\lambda, \mu)}{\rho_{6}(\lambda, \mu)} &=&
\mathrm{e}^{h}\gamma(\lambda)\alpha(\mu)-
\mathrm{e}^{-h}\alpha(\lambda)\gamma(\mu),  \label{eqn:rho4'} \\
-\frac{\rho_{10}(\lambda, \mu)}{\rho_{6}(\lambda, \mu)} &=&
-\mathrm{e}^{h}\alpha(\lambda)\alpha(\mu)-
\mathrm{e}^{-h}\gamma(\lambda)\gamma(\mu),  \label{eqn:rho9'} \\
-\frac{\rho_{9}(\lambda, \mu)}{\rho_{6}(\lambda, \mu)} &=&
-\mathrm{e}^{h}\gamma(\lambda)\gamma(\mu)-
\mathrm{e}^{-h}\alpha(\lambda)\alpha(\mu),  \label{eqn:rho10'} \\
\frac{\rho_{5}(\lambda, \mu)}{\rho_{6}(\lambda, \mu)} &=&
\frac{-\mathrm{e}^{h}\alpha(\lambda)\gamma(\mu)-
\mathrm{e}^{-h}\gamma(\lambda)
\alpha(\mu)}{\alpha^{2}(\lambda)-\alpha^{2}(\mu)},
\label{eqn:rho3'} \\
\frac{\rho_{3}(\lambda, \mu)}{\rho_{6}(\lambda, \mu)} &=&
\frac{-\mathrm{e}^{h}\gamma(\lambda)\alpha(\mu)-
\mathrm{e}^{-h}\alpha(\lambda)
\gamma(\mu)}{\alpha^{2}(\lambda)-\alpha^{2}(\mu)},
\label{eqn:rho5'} \\
-\frac{\rho_{2}(\lambda, \mu)}{\rho_{6}(\lambda, \mu)} &=&
\frac{\mathrm{e}^{-l} [ \mathrm{e}^{h}\alpha(\lambda)\gamma(\lambda)+
\mathrm{e}^{-h}\alpha(\mu)
\gamma(\mu)] }{\alpha^{2}(\lambda)-\alpha^{2}(\mu)},
\label{eqn:rho6'}
\end{eqnarray}
which are shown by direct calculation.
The relations (\ref{eqn:rhorel1})-(\ref{eqn:rhorel3}) are 
invariant under $\Phi$.

\section{Asymptotic Expansion of the Elements of the Monodromy Matrix}
\label{appendix:B}
\setcounter{equation}{0}

We shall explain below details of expansion of 
the monodromy matrix $\tilde{{\cal T}}(\lambda)$ 
in terms of $v(\lambda)^{-1}$. 
$\tilde{\cal T}_{mn}(\lambda)$ satisfies the recursion relation
\begin{equation}
\tilde{\cal T}_{m+1,n}(\lambda)=\tilde{\cal L}_{m}(\lambda)
\tilde{\cal T}_{m,n}(\lambda), \makebox[2em]{} 
\tilde{\cal T}_{m,m}(\lambda) =1,
\label{eqn:recT}
\end{equation}
where
\begin{eqnarray}
& & \tilde{\cal L}_{m}(\lambda) =  V(\lambda)^{-m-1}
{\cal L}_{m}(\lambda)V(\lambda)^{m} \nonumber \\
&& \nonumber \\
& & =
\left(
\begin{array}{cc}
(\mathrm{i}\cot {\lambda} )^{n_{m\uparrow}+n_{m\downarrow}}
&(\mathrm{i}\cot {\lambda} )^{n_{m\uparrow}}c_{m\downarrow}
\frac{\mathrm{e}^{-h(\lambda)}}{\sin \lambda} \mathrm{e}^{\mathrm{i}
mp(\lambda)}
\\
-\mathrm{i}
(\mathrm{i}\cot {\lambda} )^{n_{m\uparrow}}c^{\dagger}_{m\downarrow}
\frac{\mathrm{e}^{h(\lambda)}}{\cos \lambda} 
\mathrm{e}^{-\mathrm{i}mp(\lambda)}
&(\mathrm{i}\cot {\lambda} )^{n_{m\uparrow}-n_{m\downarrow}}
\\
c^{\dagger}_{m\uparrow}
(\mathrm{i}\cot {\lambda} )^{n_{m\downarrow}}
\frac{\mathrm{e}^{h(\lambda)}}{\cos \lambda} 
\mathrm{e}^{-\mathrm{i}mp(\lambda)}
&c_{m\uparrow}^{\dagger}c_{m\downarrow}\frac{1}{\sin \lambda \cos \lambda}
\\
\mathrm{i} c_{m\uparrow}^{\dagger}c_{m\downarrow}^{\dagger}
\frac{1}{\cos^{2} \lambda}
\tan^{2m} \lambda 
&-c_{m\uparrow}^{\dagger}(\mathrm{i}\cot \lambda)^{-n_{m\downarrow}}
\frac{\mathrm{e}^{-h(\lambda)}}{\cos \lambda} 
\mathrm{e}^{-\mathrm{i}mk(\lambda)}
\end{array}
\right.  \nonumber \\
& & \makebox[7em]{}
\left.
\begin{array}{cc}
-\mathrm{i}c_{m\uparrow}(\mathrm{i}\cot \lambda)^{n_{m\downarrow}}
\frac{\mathrm{e}^{-h(\lambda)}}{\sin \lambda} 
\mathrm{e}^{\mathrm{i}mp(\lambda)}
& -\mathrm{i} c_{m\uparrow}c_{m\downarrow}\frac{1}{\sin^{2} \lambda}
\cot^{2m} \lambda 
\\
c_{m\uparrow}c_{m\downarrow}^{\dagger}\frac{1}{\sin \lambda \cos \lambda}
&\mathrm{i}c_{m\uparrow}(\mathrm{i}\cot \lambda)^{-n_{m\downarrow}}
\frac{\mathrm{e}^{h(\lambda)}}{\sin \lambda} 
\mathrm{e}^{\mathrm{i}mk(\lambda)}
\\
(\mathrm{i}\cot {\lambda} )^{-n_{m\uparrow}+n_{m\downarrow}}
&(\mathrm{i}\cot \lambda)^{-n_{m\uparrow}}c_{m\downarrow}
\frac{\mathrm{e}^{h(\lambda)}}{\sin \lambda} 
\mathrm{e}^{\mathrm{i}mk(\lambda)}
\\
-\mathrm{i}(\mathrm{i}\cot {\lambda} )^{-n_{m\uparrow}}
c_{m\downarrow}^{\dagger}
\frac{\mathrm{e}^{-h(\lambda)}}{\cos \lambda} 
\mathrm{e}^{-\mathrm{i}mk(\lambda)}
&(\mathrm{i}\cot {\lambda} )^{-n_{m\uparrow}-n_{m\downarrow}}
\end{array}
\right). \makebox[2em]{}
\end{eqnarray}
Here we introduced new functions 
\begin{equation}
\mathrm{e}^{\mathrm{i}k(\lambda)}=
-\mathrm{e}^{2h(\lambda)}\cot \lambda, 
\makebox[1em]{}
\mathrm{e}^{\mathrm{i}p(\lambda)}=
-\mathrm{e}^{-2h(\lambda)}\cot \lambda, 
\label{eqn:defkp}
\end{equation}
which we adopted from the recent analytic Bethe Ansatz for the Hubbard 
model by Yue and Deguchi~\cite{YuDe96}.
It follows from (\ref{eqn:defv}) and (\ref{eqn:defkp})that 
\begin{equation}
\sin k(\lambda)=-\frac{v(\lambda)}{2}+\frac{\mathrm{i}U}{4}, \makebox[1em]{}
\sin p(\lambda)=-\frac{v(\lambda)}{2}-\frac{\mathrm{i}U}{4}.
\end{equation}
In the limit $|m|\rightarrow \infty$,
the above matrix $\tilde{\cal L}_{m}(\lambda)$ 
converges in the weak sense 
to the identity matrix.
Solving (\ref{eqn:recT}) iteratively we obtain 
$\tilde{\cal T}(\lambda)$ as
\begin{eqnarray}
\tilde{\cal T}(\lambda)&=&\cdots\tilde{\cal L}_{m+1}(\lambda)
\tilde{\cal L}_{m}(\lambda)
\tilde{\cal L}_{m-1}(\lambda)\cdots \nonumber \\
&=& 1+\sum_{k}(\tilde{\cal L}_{k}(\lambda)-1)
+\sum_{k>l}(\tilde{\cal L}_{k}(\lambda)-1)
(\tilde{\cal L}_{l}(\lambda)-1)+\cdots.
\label{eqn:expandtildeT}
\end{eqnarray}
To expand $\tilde{\cal L}_{m}(\lambda)$ in terms of $v(\lambda)^{-1}$, 
we consider the limit 
$\mathrm{Im}(\lambda)\rightarrow \infty$ 
and, to begin with, expand each function in terms of 
$\mathrm{e}^{2\mathrm{i}\lambda}$.
For $\mathrm{e}^{-2h(\lambda)}$ there are two possible choices of branch, 
\begin{equation}
\mathrm{e}^{-2h(\lambda)}=-\frac{U}{4}\sin 2\lambda \pm 
\sqrt{1+\left( \frac{U}{4}\sin 2\lambda \right)^{2} }.
\label{eqn:e2h}
\end{equation}
To achieve convergence of the matrix elements 
$\tilde{\cal T}_{\alpha\beta} \; (\alpha,\beta=2,3)$
we have to take the lower sign in (\ref{eqn:e2h}).
For this choice we get
\begin{eqnarray}
&&\mathrm{e}^{2h(\lambda)}=
\frac{4\mathrm{i}}{U} \mathrm{e}^{2\mathrm{i}\lambda}
+O(\mathrm{e}^{6\mathrm{i}\lambda}), \makebox[1em]{}
\mathrm{e}^{\mathrm{i}k(\lambda)}=
\frac{-4}{U} \{ \mathrm{e}^{2\mathrm{i}\lambda}
+2\mathrm{e}^{4\mathrm{i}\lambda}+
O(\mathrm{e}^{6\mathrm{i}\lambda}) \}, \nonumber \\
&&\mathrm{e}^{-\mathrm{i}p(\lambda)}=
\frac{4}{U} \{ \mathrm{e}^{2\mathrm{i}\lambda}
-2\mathrm{e}^{4\mathrm{i}\lambda}+O(\mathrm{e}^{6\mathrm{i}\lambda}) \}, 
\makebox[1em]{}
\frac{1}{v(\lambda)}=\frac{-4\mathrm{i}}{U}
\{ \mathrm{e}^{2\mathrm{i}\lambda}+
O(\mathrm{e}^{6\mathrm{i}\lambda}) \}.
\label{eqn:expandfunctions}
\end{eqnarray}
Now the leading terms in the sums in (\ref{eqn:expandtildeT})
are of order $\mathrm{e}^{2\mathrm{i}\lambda}$, 
$\mathrm{e}^{4\mathrm{i}\lambda}, \cdots$. 
Thus, 
from the first two sums in (\ref{eqn:expandtildeT}),
we get the expansion of the matrix $A(\lambda)$ up to order 
${\mathrm{e}}^{4{\mathrm{i}}\lambda}$.
Then the last equation in (\ref{eqn:expandfunctions})
yields the required expansion in $(v(\lambda))^{-1}$ up to 
second order.


\begin{thebibliography}{99}
\bibitem{EKS}
F.~H.~L.~E{\ss}ler, V.~E.~Korepin and K.~Schoutens, Phys. Rev. Lett.
67 (1991) 3848; Nucl. Phys. 
B372 (1992) 599;
{\it ibid.} 
B384 (1992) 431.
\bibitem{HeLi71}
O.~J.~Heilmann and E.~H.~Lieb, Ann. NY Acad. Sci. 172 (1971) 583.
\bibitem{Ya89}
C.~N.~Yang, Phys. Rev. Lett. 63 (1989)2144;
C.~N.~Yang and S.~C.~Zhang, Mod. Phys. Lett. B4 (1990) 759.
\bibitem{LiWu68}
E.~H.~Lieb and F.~Y.~Wu, Phys. Rev. Lett. 20 (1968) 1445.
\bibitem{Hubbardbook}
V.~E.~Korepin and F.~H.~L.~E{\ss}ler, eds, Exactly Solvable 
Models of Strongly Correlated Electrons (World Scientific, Singapore, 
1994).
\bibitem{Ya67}
C.~N.~Yang,  Phys. Rev. Lett.  19 (1967) 1312.
\bibitem{Ga67}
M.~Gaudin,  Phys. Lett.  A24 (1967) 55.
\bibitem{Sha86}
B.~S.~Shastry, Phys. Rev. Lett. 56 (1986) 1529;
{\it ibid.} 56 (1986) 2453.
\bibitem{Sha88}
B.~S.~Shastry, J. Stat. Phys. 50 (1988) 57.
\bibitem{WaOlAk87}
M.~Wadati, E.~Olmedilla and Y.~Akutsu, J. Phys. Soc. Jpn. 
56 (1987) 1340.
\bibitem{OlWaAk87}
E.~Olmedilla, M.~Wadati and Y.~Akutsu, J. Phys. Soc. Jpn. 
56 (1987) 2298.
\bibitem{OlWa88}
E.~Olmedilla and M.~Wadati, Phys. Rev. Lett. 
60 (1988) 1595.
\bibitem{ShiWa95}
M.~Shiroishi and M.~Wadati, J. Phys. Soc. Jpn. 
64 (1995) 57.
\bibitem{RaMa96}
P.~B.~Ramos and M.~J.~Martins, preprint(1996) (hep-th/9605141).
\bibitem{UgKo94}
D.~B.~Uglov and V.~E.~Korepin, Phys. Lett. A190 (1994) 238.  
\bibitem{Dri86}
V.~G.~Drinfeld, in  Proc.  of ICM  
(Berkeley, California, USA, 1986) 269.
\bibitem{ChaPre}
V.~Chari and A.~N.~Pressley,  L'Enseignement Math\'ematique 
 36 (1990) 267; 
J. Reine. Angew. Math.  417 (1991) 87.
\bibitem{LeSmi92}
A.~LeClair and F.~A.~Smirnov,  Int. J. Mod. Phys.  A7 (1992)
2997.
\bibitem{EsKo94}
F.~H.~L.~E{\ss}ler and V.~E.~Korepin, Phys. Rev. Lett. 72 (1994) 908;
Nucl. Phys. B426 (1994) 505.
\bibitem{JMMN}
M.~Jimbo, K.~Miki, T.~Miwa and A.~Nakayashiki, Phys. Lett. A168 
(1992) 256.
\bibitem{BaGe95}
F.~Gebhard and A.~E.~Ruckenstein, Phys. Rev. Lett. 68 (1992) 244;
P.~A.~Bares and F.~Gebhard, Europhys. Lett. 29 (1995) 573.
\bibitem{GoIn96}
F.~G\"ohmann and V.~Inozemtsev, Phys. Lett. A214 (1996) 161. 
\bibitem{MuWa96a}
S.~Murakami and M.~Wadati, J. Phys. Soc. Jpn. 65 
(1996) 1227.
\bibitem{MuWa96b}
S.~Murakami and M.~Wadati, Connection between Yangian Symmetry and the 
Quantum Inverse Scattering Method, preprint (1996).
\bibitem{FaTa}
L.~D.~Faddeev and L.~A.~Takhtajan, Hamiltonian Methods in the 
Theory of Solitons (Springer-Verlag, Berlin, 1987).
\bibitem{Fa80}
L.~D.~Faddeev, Sov. Sci. Rev. Math. Phys. C1 (1980) 107.
\bibitem{Tha81}
H.~B.~Thacker, Rev. Mod. Phys. 53 (1981) 253.
\bibitem{GoMu96}
F.~G\"ohmann and S.~Murakami, Algebraic and Analytic Properties of the 
One-dimensional Hubbard Model, preprint (1996).
\bibitem{Hal94}
F.~D.~M.~Haldane, in Proc. of the 16th Taniguchi Symposium on 
Condensed Matter Physics (Springer, Berlin-Heidelberg-New York, 1994).
\bibitem{IzKo81}
A.~G.~Izergin and V.~E.~Korepin, Sov. Phys. Dokl. 26 (1981) 653.
\bibitem{KBIbook}
V.~E.~Korepin, N.~M.~Bogoliubov and A.~G.~Izergin, 
{\it Quantum Inverse Scattering Method and Correlation Functions} 
(Cambridge University Press, Cambridge, 1993) p.160.
\bibitem{Gro}
H.~Grosse, Lett. Math. Phys. 18 (1989) 151.
\bibitem{Dunkl}
A.~Polychronakos, Phys. Rev. Lett. 69 (1992) 703;
J.~A.~Minahan and A.~Polychronakos, Phys. Lett. B302 (1993) 265;
D.~Bernard, M.~Gaudin, F.~D.~M.~Haldane and V.~Pasquier,  J. Phys. 
 A26 (1993) 5219.
\bibitem{Ug95}
D.~B.~Uglov, Phys. Lett. A199 (1995) 353.  
\bibitem{HiMu96}
K.~Hikami and S.~Murakami, Phys. Lett. A221 (1996) 109.
\bibitem{YuDe96}
R.~Yue and T.~Deguchi, preprint(1996) (cond-mat/9606039).
\end{thebibliography}
\end{document}